\documentclass[prb,a4paper,showpacs,groupedaddress,floatfix,superscriptaddress,twocolumn]{revtex4}
\usepackage[english]{babel}
\usepackage{amsmath,amssymb}
\usepackage{dsfont}
\usepackage{txfonts}
\usepackage{upgreek}
\usepackage{subfigure}
\usepackage{graphicx}
\usepackage{epstopdf}
\usepackage{dcolumn}
\usepackage{float}
\usepackage{color}

\usepackage{bm}
\usepackage{epstopdf}

\usepackage{hyperref}

\newcommand{\varF}{{\mathcal{F}}}
\newcommand{\varH}{{\mathcal{H}}}
\newcommand{\varU}{{\mathcal{U}}}

\newcommand{\mrm}[1]{\mathrm{#1}}

\newcommand{\sgn}{\mathop{\operator@font sgn}}

\newcommand{\ket}[1]{\left|#1\right\rangle}
\newcommand{\bket}[2]{\langle #1|#2\rangle}

\def\bra#1{\mathinner{\langle{#1}|}} 
\def\ket#1{\mathinner{|{#1}\rangle}}

\begin{document}

\preprint{APS/123-QED}

\title{Ballistic spin resonance in multisubband quantum wires}

\author{Marco O. Hachiya}
\affiliation{Instituto de F\'{\i}sica de S\~ao Carlos, Universidade de S\~ao Paulo, 13560-970 S\~ao Carlos, S\~ao Paulo, Brazil}
\author{Gonzalo Usaj}
\affiliation{ Centro At\'omico Bariloche and Instituto Balseiro, Comisi\'on Nacional de Energ\'ia At\'omica, 8400 San Carlos de Bariloche, Argentina}
\affiliation{Consejo Nacional de Investigaciones Cient\'ificas y T\'ecnicas (CONICET), Argentina}
\author{J. Carlos Egues}
\affiliation{Instituto de F\'{\i}sica de S\~ao Carlos, Universidade de S\~ao Paulo, 13560-970 S\~ao Carlos, S\~ao Paulo, Brazil}

\date{\today}

\begin{abstract}
Ballistic spin resonance was experimentally observed in a
quasi-one-dimensional wire by Frolov \textit{et al.} [Nature (London) {\bf 458}, 868 (2009)]. The spin resonance was
generated by a combination of an external static magnetic field and the
oscillating effective spin-orbit magnetic field due to periodic 
bouncings of the electrons off the boundaries of a narrow channel. An increase of the D'yakonov-Perel spin relaxation rate was observed when the frequency of
the spin-orbit field matched that of the Larmor precession frequency around
the external magnetic field. Here we develop a model to account for the
D'yakonov-Perel mechanism in multisubband quantum wires with both the Rashba
and Dresselhaus spin-orbit interactions. Considering elastic spin-conserving impurity scatterings in the time-evolution operator (Heisenberg representation), we extract the spin relaxation time by evaluating the time-dependent expectation value of the spin operators. 
The magnetic field dependence of the nonlocal voltage, which is related to the spin relaxation time behavior, shows a wide plateau, in agreement with the experimental observation. This plateau arises due to injection in higher subbands and small-angle scattering. 
In this quantum mechanical approach, the spin resonance occurs near
the spin-orbit induced energy anticrossings of the quantum wire subbands
with opposite spins. 
We also predict anomalous dips in the spin relaxation time as a function of the magnetic field in systems with strong spin-orbit couplings.
\end{abstract}

\pacs{72.25.Rb, 72.25.-b, 73.21.Hb}

\maketitle

\section{\label{sec:intro} Introduction}
The spin-orbit (SO) coupling is an essential ingredient to control and manipulate the spin degree of freedom in potential spintronic devices. 
In zinc-blend based quantum wells, the SO induced momentum-dependent spin-splitting is caused by structural and bulk inversion asymmetry, respectively, leading to the Rashba and Dresselhaus SO interactions. 
In particular, the Rashba SO strength can be tuned via external gates \cite{Nitta, Engels97} allowing controlled coherent spin rotations in a quasi-one dimensional channel with ferromagnetic source and drain contacts. This is the well-known Datta-Das spin-FET proposal\cite{Dattadas, EguesAPL}. 
Despite providing a way to control and manipulate the electron spin, the SO coupling also plays a crucial role in the spin relaxation in dimensionally-constrained semiconductor nanostructures. 

Regarding the spin relaxation in quantum wires, the main mechanism in zinc-blende based nanostructures involves the SO interaction combined with random multiple scattering events. Both processes combined are responsible for misalignment of an ensemble of initially polarized spins, a process known as the D'yakonov-Perel (DP) relaxation mechanism\cite{DP72}. This mechanism is directly connected to the fact that the SO interaction can be described by a momentum-dependent effective magnetic field.
Thus scattering events will randomize the electron momentum direction generating a random fluctuating SO magnetic field causing spin relaxation. 
The DP spin relaxation time is inversely proportional 
to the momentum scattering time leading to its increasing as the channel width becomes comparable to the electron mean free path\cite{Kiselev04, Malshukov}.

\begin{figure}[htb]
\includegraphics[width=.5\textwidth]{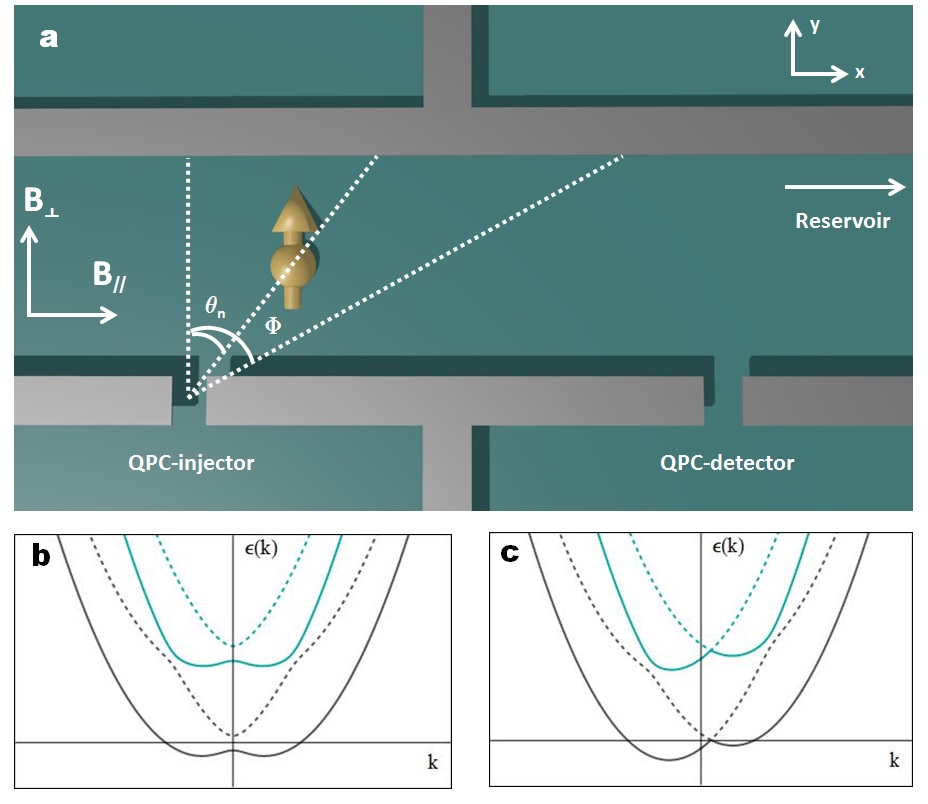}  
\caption{(a) Schematic view of a quasi-one dimensional channel formed in a 2D electron gas.
A spin-polarized current is injected in a multisubband quantum wire via a spin-selective QPC with an angular spread $\Phi$. 
The spin current diffuses towards a spin-unpolarized reservoir.
Each electron is assumed to be uniformly distributed in the subbands with quantum number $n$ such that $\theta_n < \Phi$. After undergoing multiple random scatterings, the ensemble spin polarization will decay as a consequence of electron spins precessing around distinct fluctuating momentum-dependent effective magnetic fields due to the SO interaction (D'yakonov-Perel mechanism). 
Energy spectrum of a quantum wire with SO interaction and an external magnetic field (b) parallel $B_{\parallel}$  and (c) perpendicular $B_{\perp}$ to the quantum wire. The former case opens a gap at $k=0$ and the latter case induces an asymmetry of the energy branches depending on the sign of $k$. 
The subband-spin mixing term $\varH_{\mrm{SO}}^{\parallel} = i (\alpha - \beta) \partial_y \sigma_x$ induces energy anticrossings of the quantum wire subbands with opposite spins.
In the absence of $\varH_{\mrm{SO}}^{\parallel}$, the magnetic field-tunable level crossing defines the resonance condition for the BSR. }
\label{fig:plot0BSR}
\end{figure}

\begin{figure}
\includegraphics[width=.47\textwidth]{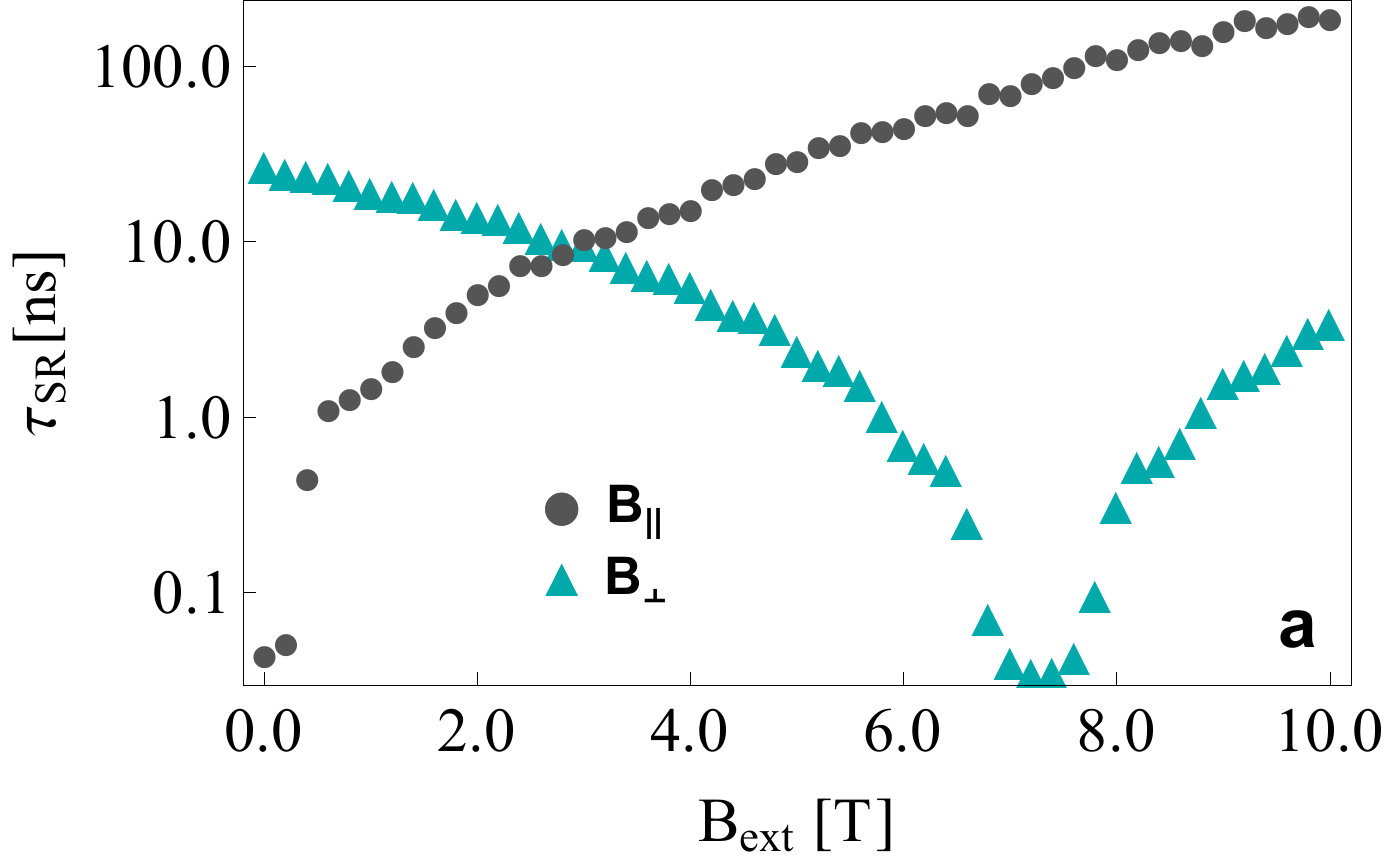}
\includegraphics[width=.46\textwidth]{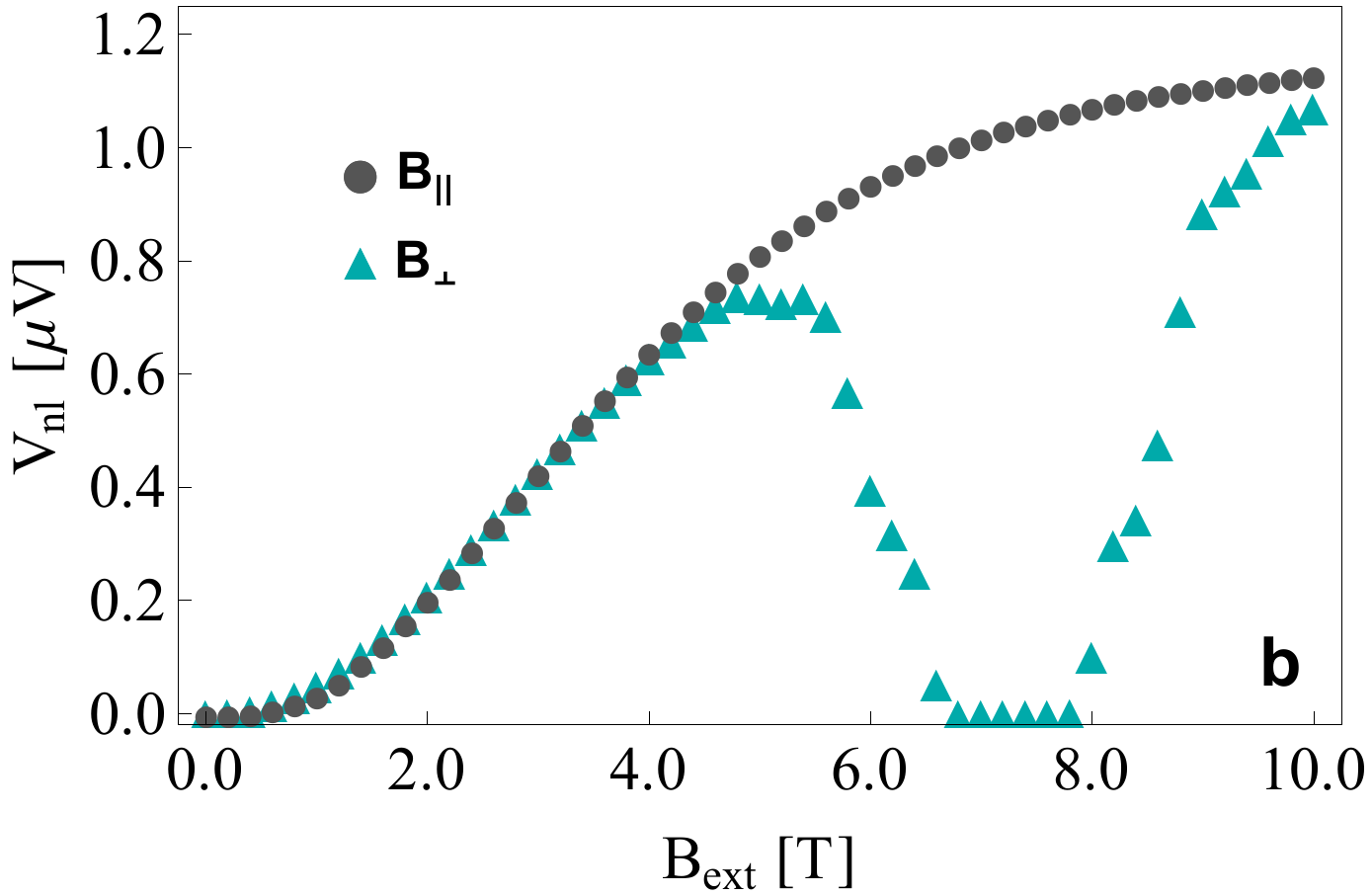}
\includegraphics[width=.46\textwidth]{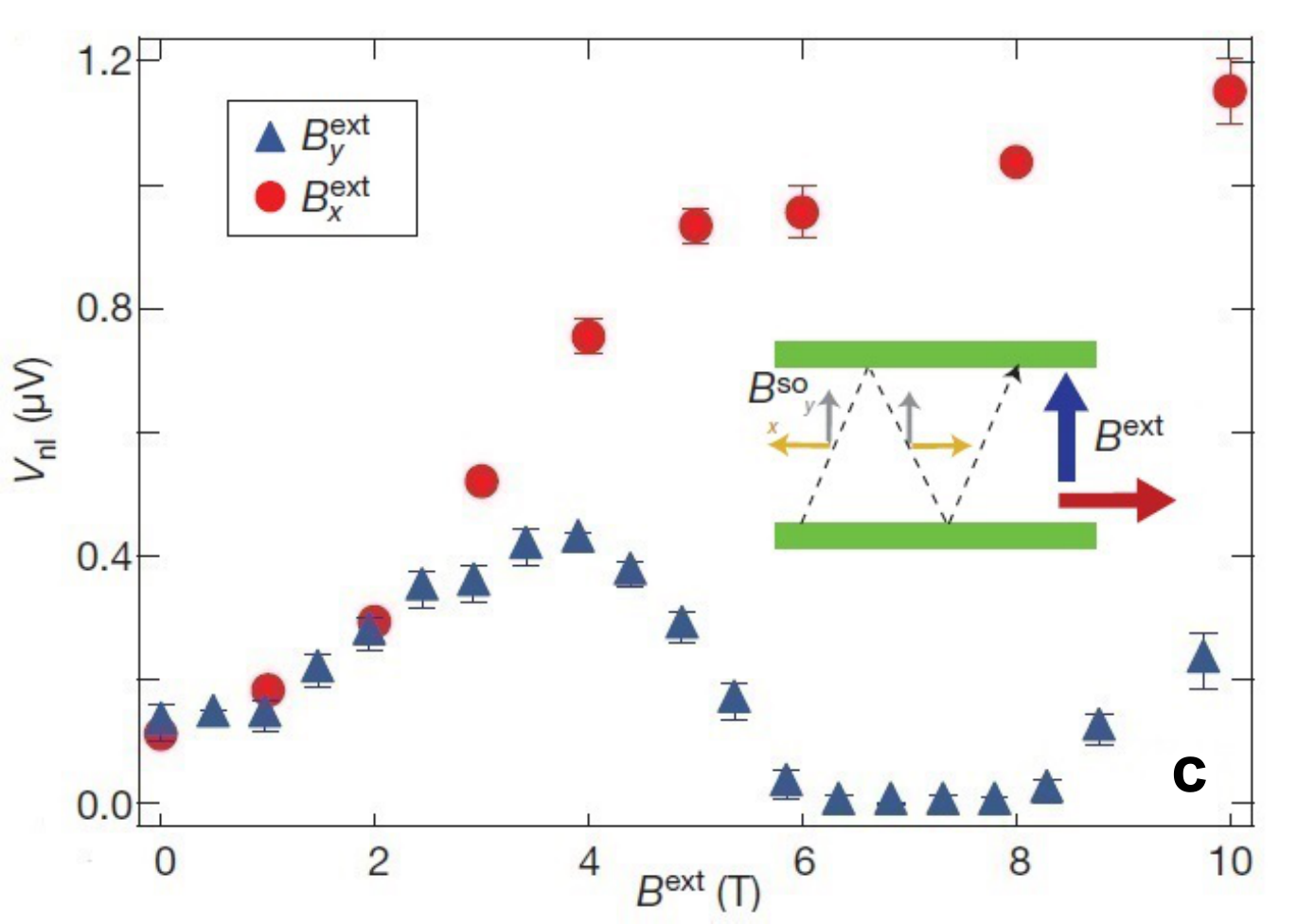}
\caption{(a) Dependence of the spin relaxation time $\tau_{\mrm{SR}}$ [(b) Nonlocal voltage $V_{\mrm{nl}}$] on the external magnetic field $\mathbf{B}_{\mrm{ext}}$
For $B^{\perp}$ (light triangles), a clear dip of the spin relaxation time emerges near the SO-induced energy anticrossings of the quantum wire subbands with opposite spins. The resonance condition, given by Eq.~(\ref{eq:resonance}), is fulfilled for $B^{\perp} \approx 8$ T for the highest subband $n_j=17$ with $\Lambda=1$, while $\tau_{\mrm{SR}}$ and consequently $V_{\mrm{nl}}$ increase monotonically with $B^{\parallel}$ (dark circles).  
(c) Data extracted from the BSR experiment \cite{Bsr} show the same behavior when compared with our numerical results including a wide resonance plateau.
We have used the following parameters in our simulation: $\delta t = 2$ $\mrm{ps}$, $N = 1000$, $\Delta = 3^{\circ}$, $L=1$ $\mu m$, $\Phi =30^{\circ}$, number of electrons considered in the ensemble $N_{\mrm{ens}}=1000$, electronic density $n_{1D} \approx 10^{8}$ $m^{-1}$.
For the nonlocal voltage $V_{\mrm{nl}}$ we used the channel resistivity $\rho=40$ $\Omega$, left (right) end of the channel $L_l = 30$ $\mu m$ ($L_r = 70$ $\mu m $), position of QPC injector (detector) $x_{\mrm{inj}}=0$ ($x_{\mrm{det}}= 20$ $\mu m$), temperature $T =300$ mK.
For the GaAs quantum well, $|(\alpha + \beta)|=0.05$ $\mrm{meV.nm}$, $|(\alpha - \beta)|=0.2$ $\mrm{meV.nm}$ \cite{calsaverini}, $\left|g\right|=0.44$\cite{vurga} and $m=0.067$ $m_0$\cite{vurga}, where $m_0$ is the bare electron mass. }
\label{fig:plot1BSR}
\end{figure}

In general, the spin relaxation time is a monotonic function of the external magnetic field\cite{OpticalOrientation, DyakonovBook, Frolov09}.
Nevertheless, a nonmonotonic behavior can arise by combining an external time-independent magnetic field and periodic oscillations of the SO effective magnetic field -- in another words, an electron spin resonance\cite{Duckheim} in the absence of the {\it external} oscillating fields, namely ballistic spin resonance (BSR).  
In a semiclassical picture, BSR could be interpreted considering an electron injected by a spin-polarized quantum point contact (QPC) traveling along a ballistic channel towards a large spin-unpolarized reservoir. 
Each electron experiences random scattering events as well as periodic bouncings off the lateral confinement \cite{Luscher10, Kiselev04, Koop}. 
The resonance condition is achieved matching the frequency of the SO field with the Larmor precession frequency around the external magnetic field, the spin-flip probability is maximized thus increasing the spin relaxation rate. Then, the randomized electron spin can be detected using another spin-selective QPC.
A nonlocal voltage, measured between the detector QPC and the reservoir\cite{Bsr, Frolov09}, quantifies the spin accumulation along the channel and it is suppressed whenever the resonance condition is fulfilled.

In the present work, we introduce a model to account for the DP mechanism in multisubband quantum wires, Figs. 1(a)-1(c). 
We monitor the spin dynamics for an ensemble of electrons undergoing random scattering events transitioning among quantum wire subbands. Averaging the spin dynamics over an ensemble, we are able to extract the spin relaxation time.
We study the dependence of the spin relaxation time on the external magnetic field perpendicular to the wire and the emergence of a nonmonotonic behavior
characterizing the ballistic spin resonance. 
Within our model, the spin resonance occurs at the quantum wire subband anticrossing induced by the SO interaction.
Each electron in the ensemble is redistributed due to scattering mechanisms among different subbands, since each subband has a resonance condition for distinct values of the external magnetic field leading to an enlargement of the BSR dip into a wide plateau.
On the other hand, the spin relaxation time presents a monotonic behavior when the magnetic field is aligned to the wire and consequently to the oscillating SO field.
Our theoretical results present (see Fig.~\ref{fig:plot1BSR}) the same behavior for the spin relaxation time as a function of the magnetic field in both directions of the external magnetic field as shown in the BSR experiment\cite{Bsr}. 
Nevertheless, we also predict the presence of anomalous BSR dips in the spin relaxation time as a function of the magnetic field even when it is aligned with the wire orientation. 
We predict that the nonmonotonic behavior could be experimentally observed in systems with strong SO couplings. In this case, a strong component of the SO magnetic field can tilt the spin perpendicularly to the oscillating field also quickening the spin relaxation rate. 
This paper is organized as follows: In Sec.~\ref{sec:model}, we describe our model.
In Sec.~\ref{sec:bsr}, we present the numerical results of the magnetic field dependence of the spin relaxation time.
In Sec.~\ref{sec:anomalousbsr}, we predict and discuss the presence of anomalous BSR dips.
We conclude in Sec.~\ref{sec:conclusion}; we present the conclusion and discussions about the potential applications of the model such as investigating the width dependence and anisotropy of the spin relaxation time.

\section{\label{sec:model} The Model}

Consider a high-mobility 2D electron gas formed in a zinc-blende semiconductor crystal. 
The {\it linear-in-p} Rashba and Dresselhaus SO coupling\cite{RashbaSO, DresselhausSO, cubicDress} can be represented by a momentum-dependent effective magnetic field

\begin{equation}
\varH_{\mrm{SO}}=\frac{1}{2}g \mu_B   \mathbf{B}_{\mrm{SO}} \cdot \bm{\sigma}, 
\quad \mathbf{B}_{\mrm{SO}} = \frac{2}{g \mu_B \hbar} \left[
\begin{array}{c}
\quad (\alpha - \beta) p_y \\
-(\alpha + \beta) p_x \\
\end{array} \right]
\label{eq:BSO}.
\end{equation}
for a coordinate system such that  $x||$[110], $y||$[$\overline{1}$10].
Here, $\alpha$ and $\beta$ correspond to the Rashba and Dresselhaus SO coupling strengths, respectively. 
Also, $\mathbf{p}$ denotes the electron momentum, $\boldsymbol\sigma$ the Pauli matrices.
A multisubband quantum wire can be engineered in this system by parallel spatially separated metal gates (split gate) on top of a 2D electron gas. 
Thus, the electrostactic potential depletes the electrons under the gates forming a quasi-one-dimensional channel for the conduction electrons.
A proper geometry for the split gate allows a pure spin current injection via a spin-selective quantum point contact (QPC). 
Similarly, the corresponding spin accumulation due to this spin current can be detected using a spatially separated QPC\cite{Bsr}.
Considering a square wire confinement with width $L$, the Hamiltonian describing the system reads

\begin{eqnarray}
\varH=\frac{{\bf p}^2}{2m}+\frac{1}{2}g \mu_B \left(  \mathbf{B}_{\mrm{SO}} +   \mathbf{B}_{\mrm{ext}}  \right) \cdot \boldsymbol\sigma  + V(y),
\label{eq:Hsquare}
\end{eqnarray}
with effective mass $m$, $V(y)=0$ for $0 \leq y \leq L$ and $V(y) \rightarrow \infty$ elsewhere. The external magnetic field ${\bf B}_{\mrm{ext}}$ applied in the plane of the 2D electron gas has two purposes:
it defines the spin polarization of the electron injected in the quantum wire through a QPC and it serves as a controllable external knob for the spin resonance condition.
In order to determine the electron spin dynamics, we have to obtain the eigenenergies and eigenstates of the Hamiltonian $\varH$ which describes our system.
This can be achieved numerically for a given $k$ by projecting the Hamiltonian $\varH$ in a truncated subband-spin Hilbert space $\varF = \left\{ \ket{n,k,s_i}; n={1,2,...n_{T}},k, s_i=\uparrow_i,\downarrow_i  \right\}$, where $i=x,y,z$, $n_{T}$ is the total number of subbands in the subspace $\varF$ and $s_i$ denotes the spin component along the $i$ direction. 
Here, $k$ represents the wavevector of the plane wave solution along the quantum wire and $n$ is the quantum number related with transverse direction of the quantum wire, i.e., $\bket{\mathbf{r}}{n,k,s_i} =\sqrt{2/L}\sin(n \pi y /L) e^{i k x} \chi_i $, where $\chi_i$ is the spinor in the $\sigma_i$ basis. 


Consider an electron injected initially into the subband labeled $n_{j}$ of this quantum wire.
Its quantum dynamics is entirely described by the time-evolution operator $\varU(k,t) = \mrm{exp} \left[-(i/\hbar)\varH( k) t\right]$.
Thus the electron spin dynamics of the $i=x,y$ component initially injected in a general state $\ket{n_j,k,s_i}$, with the spin projection axis aligned with $\mathbf{B}_{\mrm{ext}}$, is obtained by numerically calculating the time-dependent expectation values of the respective Pauli spin matrix $ \bar{\sigma}_i(t) = \bra{n_j,k,s_i} \varU^{\dagger}(k,t) \sigma_i \varU(k,t) \ket{n_j,k,s_i}=\bra{n_j,k,s_i} \sigma_i(t)\ket{n_j,k,s_i}$ in the Heisenberg representation \cite{Zitterprl, Zitterprb}. 
More explicitly, we have

\begin{eqnarray}
\bar{\sigma}_i(t) = \bra{n_{j},k,s_i} \mathsf{P}_k \tilde{\varU}^{\dagger}(k,t) \mathsf{P}_k^{-1} \sigma_i \mathsf{P}_k \tilde{\varU}(k,t) \mathsf{P}_k^{-1} \ket{n_{j},k,s_i},
\label{eq:avg}
\end{eqnarray}
where $\mathsf{P}_k$ is a matrix whose columns are composed of the eigenvector components which diagonalize the Hamiltonian $\varH$ for a given $k$. Here, we have used the similarity transformation $\tilde{\varU}(k,t) = \mathsf{P}_k^{-1} \varU(k,t) \mathsf{P}_k$\cite{eigendecomp}, where $\tilde{\varU}(k,t)$ assumes a diagonal form.

{\it Scattering mechanisms}. The preceding approach to calculate the electron spin dynamics\cite{Zitterprl, Zitterprb} can be generalized to include multiple random scattering events.  
Here, we consider wave packets propagating freely between collisions.
We allow for transitions between quantum wire subbands after each scattering.
Between these transitions, the electron spin will precess around the SO and external magnetic fields. 
This characterizes the DP mechanism in multisubband quantum wires. 
%
%
Here, we consider large-angle and small-angle scatterings which suffice to describe the experimental data.
The large-angle scattering mechanism is taken into account considering that an elastic spin-conserving impurity scattering occurs with a probability $\delta t/ \tau$ for a time interval $\delta t$, where $\tau$ is the mean-free time. 
After each scattering, the electron momentum orientation is randomized. It can make transitions to all equally probable subbands at the Fermi energy representing a large angle scattering.
A ballistic quantum wire is assumed such that the mean-free path $\lambda$ is much larger than the quantum wire width, $\lambda \gg  L$.
Another significant source of scattering is the ionized donors responsible for initially forming the 2D electron gas. These dopants are spatially separated from the electron gas. So, electrons feel a weaker screened Coulomb potential
leading to a majority of small-angle scattering events, and rarely a full backscattering.
This scattering mechanism is implemented choosing a random number $\tilde{\Theta}$ from a normal distribution with zero mean and standard deviation $\Delta$ for each timestep. We consider an electron coming from the subband $n_k$ and making a transition to the subband $n_l$ at the Fermi energy if $\Theta_{n_k,n_{l-1}} \le \tilde{\Theta}  \le \Theta_{n_k,n_{l+1}}$, where $\Theta_{n_k,n_l}=\theta_{n_k} - \theta_{n_l}$.
Here, we ascribe a set of angles ${\theta_n}$ to the electron quantum states.
For a given Fermi momentum $k_F$, the injection angle between the transverse direction and its Fermi momentum can be defined as $\theta_{n} = \arcsin (k_n/k_F)$, where $k_n = \sqrt{k_F^2-(n \pi /L)^2}$.\cite{Glazman91} 
{\it Generalized expectation value of the spin operators}. With these momentum scattering mechanisms considered, the generalized time evolution operator after $N$ scatterings for each time interval $\delta t$ is sequentially assembled as, 

\begin{eqnarray}
\mathsf{U}_N(t)&=&\mathsf{U}(\gamma_1 k_{n_{1}},\delta t)\mathsf{U}(\gamma_2 k_{n_{2}}, \delta t)...\mathsf{U}(\gamma_N k_{n_{N}},  \delta t) \nonumber\\ 
					&=& \prod_{\nu=1}^{N} \mathsf{U}(\gamma_\nu k_{n_{\zeta}}, \delta t)
\label{eq:timescattu}
\end{eqnarray}
Here, $\mathsf{U}(\gamma_\nu k_{n_{\zeta}},t)=\mathsf{P}_{\gamma_\nu k_{n_{\zeta}}} \tilde{\varU}(\gamma_\nu k_{n_{\zeta}},t) \mathsf{P}_{\gamma_\nu k_{n_{\zeta}}}^{-1}$\cite{SW} for the $\nu$-th scattering event to the subband $n_{\zeta}$, where $\gamma_\nu=\pm 1$ depending on whether the electron has scattered backwards $\gamma_\nu$=-1 or moved forward $\gamma_\nu$=+1 at the time $t=\nu \delta t$. 
We have that $n_{\zeta}$ is an integer random number, with $1 \leq  n_{\zeta} \leq n_{T}$, sorted out according to the scattering mechanisms considered, as explained in Sec.~\ref{sec:model}.
Thus considering scattering between quantum wire subbands, we have a generalization of the expectation value of the spin operator $\bar{\sigma}_i(t) = \bra{n_j,k,s_i} \mathsf{U}_N^{\dagger}(t) \sigma_i \mathsf{U}_N(t) \ket{n_j,k,s_i}$. 
This procedure can be repeated for an ensemble of initially spin-polarized electrons in order to obtain the average spin polarization as a function of time, $P_{i}(t)=\sum^{N_{\mrm{ens}}}_{\mu=1}  \bar{\sigma}_i^\mu(t) / N_{\mrm{ens}}$, where $P_{i}$ is the polarization along the $i$ direction for the $\mu$-th electron, and $N_{\mrm{ens}}$ is the total number of electrons considered in the simulation.
The noncommutativity of the time-evolution operators describing successive scatterings implies that the path followed by the electron matters in a multisubband quantum wire. 
Therefore, random paths result in random spin precession for each electron and spin relaxation for the whole ensemble
(see Appendix~\ref{app:simplemodel} for a more qualitative picture of the DP mechanism in quantum wires).
As time goes by, the average ensemble spin polarization decays exponentially with a timescale given by the spin relaxation time $\tau_{\mrm{SR}}$, i.e., $P_{i}(t)=P_{i}(t=0) e^{-t/\tau_{\mrm{SR}}}$. 
This whole procedure can be repeated for different external magnetic fields thus allowing us to extract $\tau_{\mrm{SR}} ({\bf B}_{\mrm{ext}})$ using a single-exponential decay fit. 
Notice that we consider scattering events as transitions between different quantum wire subbands. 
Since each subband will have a distinct resonance condition, we find that the corresponding resonance dip evolves into a wide plateau, in agreement with the experimental findings\cite{Bsr}.
This is in contrast with semiclassical Monte Carlo simulations\cite{Kiselev04, Luscher10} where the electron moves in a 2D electron gas
undergoing momentum randomizing scattering events and bouncing off the walls of the channel.
In this case, each BSR dip has a well-defined value for the external magnetic field and depends on the electron Fermi velocity and the channel width\cite{Bsr}.
In the next section, we will analyze the magnetic field dependence of the spin relaxation time in a realistic system.
We will compare our numerical results with the experimental features of the BSR.
%



\section{\label{sec:bsr} Ballistic spin resonance}

In order to simplify our discussion and have a better understanding of the role of each term in the Hamiltonian (\ref{eq:Hsquare}), we separate the total Hamiltonian as $\varH=\varH_0 + \varH_{\mrm{SO}}^{\parallel} + \varH_{\mrm{SO}}^{\perp} + \varH_Z^{\parallel} + \varH_Z^{\perp}$, where we define the quantum wire Hamiltonian $\varH_0 = \frac{\hbar^2 k^2 }{2m} + \frac{n^2 \hbar^2 \pi^2}{2m L^2}$, the SO contribution $\varH_{\mrm{SO}}^{\perp} = - (\alpha + \beta) k \sigma_y$, $\varH_{\mrm{SO}}^{\parallel} = i (\alpha - \beta) \partial_y \sigma_x$, and the Zeeman terms $\varH_Z^{\parallel} = g \mu_B B_x \sigma_x/2$, $\varH_Z^{\perp} = g \mu_B B_y \sigma_y/2$. 
Here, the superscripts $\perp$ and $\parallel$ denote the SO and external magnetic fields components perpendicular ($\hat{y}$) and parallel ($\hat{x}$) to the quantum wire, respectively.

According to the experimental setup used to detect the BSR\cite{Bsr}, electrons are injected using a voltage applied through a QPC (injector) and diffuses along the multisubband quantum wire until their detection by another QPC (detector). 
Both QPCs are fully spin polarized (conductances equal to $e^2/h$) with quantization axis defined by the external magnetic field $\mathbf{B}_{\mrm{ext}}$.
The pure spin-polarized current starts to relax with the characteristic time $\tau_{\mrm{SR}}$ according to the DP mechanism. 
We now analyze two cases where the electron is injected with its spin aligned to either $B^{\parallel}$ or $B^{\perp}$.

Initially, we inject an ensemble of electrons into the quantum wire with an angular spread $\Phi$ relative to the transverse direction\cite{Topinka}.
These electrons are uniformly distributed over the subbands with quantum numbers $n$ within $\Phi$\cite{footnote1}; i.e., $\theta_n < \Phi$. Notice that this requirement is fulfilled only by the higher subbands. 
Consider a particular case where an electron is injected in the subband $n_j$ near the energy anticrossing with its spin pointing along the $y$ axis aligned with $B^{\perp}$ ($B^{\parallel} = 0$). It will undergo two processes caused by $\varH_{\mrm{SO}}^{\parallel}$ in Eq.(\ref{eq:Hsquare}): an intersubband transition due to momentum operator $- i \hbar \partial_y$ connecting different orbital states and a spin-flip along the $y$ direction due to the operator $\sigma_x$, since $\bra{n_{j},k,\downarrow_y} \varH_{\mrm{SO}}^{\parallel} \propto \partial_y \sigma_x \ket{n_{j} \pm \Lambda,k,\uparrow_y} \neq 0$, where $\Lambda$ is an odd integer ($\Lambda=1,3,5...$). Thus the spin relaxation time also will strongly decrease near the energy level anticrossing induced spin-orbital mixing caused by the term $\varH_{\mrm{SO}}^{\parallel}$. At the energy anticrossing (resonance condition), the spin-flip probability is maximized thus quickening the spin relaxation process which characterizes the BSR effect.  
The resonance condition is determined by the energy-level crossings in the spectrum of $\left[ \varH_0 + \varH_{\mrm{SO}}^{\perp} + \varH_Z^{\perp}  \right]\ket{n,k,s_y} = \epsilon_{n,k,s_y} \ket{n,k,s_y}$. Thus the crossing $\epsilon_{n_j,k,\downarrow_y} =\epsilon_{n_j\pm\Lambda,k,\uparrow_y}$ occurs for the $B^{\perp}_{\mrm{BSR}}$ given by

\begin{eqnarray}
\frac{1}{2} g \mu_B B^{\perp}_{\mrm{BSR}} = \frac{\pi^2 \hbar^2}{4 m L^2} \left[ \pm 2n_{j} \Lambda + \Lambda^2    \right] + (\alpha + \beta) k_F .
\label{eq:resonance}
\end{eqnarray}
where we have used $k_F = k_F^{n_j} \approx k_F^{n_j + \Lambda}$.
The 2D semiclassical limit for this resonance condition can be obtained relating the injection subband $n_j$ with the Fermi velocity $v_F^{\perp}$ and the channel width $L$ as $v_F^{\perp} = \hbar \pi  n_j  / m L  $. Assuming that $g \mu_B B^{\perp}_{\mrm{BSR}} \gg |(\alpha + \beta)| k_F$ and $n_j \gg 1$, isolating $n_j$ and substituting into Eq.~\ref{eq:resonance}, we recover the resonance frequency $f_{\Lambda} = v_F^{\perp}/ 2 L \times \Lambda  = g \mu_B B^{\perp}_{\mrm{BSR}}/ h$, in agreement with Ref.~\onlinecite{Bsr}.  
In contrast, if the electron spin is initially aligned along the $x$ axis for $B^{\parallel}$ ($B^{\perp} = 0$), no BSR is observed.
Although $\varH_{\mrm{SO}}^{\parallel}$ can cause intersubband transition, this term is not able to flip the spin since the spin operator $\sigma_x$ is acting on its eigenstate, $\bra{n_{j},k,\uparrow_x} \varH_{\mrm{SO}}^{\parallel} \propto \partial_y \sigma_x \ket{n_{j} \pm \Lambda,k,\downarrow_x} = 0$. Thus even fulfilling the condition for the crossing of energy levels with opposite spins, 
there is no spin resonance in the quantum wire, and consequently, the spin relaxation time has a monotonic dependence with $B^{\parallel}$. 
Notice that in the weak SO coupling regime, $g \mu_B B^{\parallel} /2 \gg |(\alpha + \beta)| k_F $ where the resonance occurs according to the BSR experiment in a GaAs quantum well\cite{Bsr}. As a consequence, the SO magnetic field is not able to tilt the spin alignment from the orientation parallel to the channel. 
Two distinct behaviors then arise observing the magnetic field dependence of $\tau_{\mrm{SR}}$ (see Fig.~\ref{fig:plot1BSR}) depending on the in-plane $\mathbf{B}_{\mrm{ext}}$ orientation. For a $B^{\parallel}$, the $\tau_{\mrm{SR}}(B^{\parallel})$ increases monotonically for all values of $B^{\parallel}$. On the other hand, $\tau_{\mrm{SR}}(B^{\perp})$ is strongly suppressed around the energy anticrossing induced spin-orbit mixing.
These different behaviors can be quantified experimentally via a nonlocal voltage $V_{\mrm{nl}}$\cite{Frolov09}. This quantity is related to the variation of the chemical potential from the detector QPC to a large spin-unpolarized reservoir [see Fig.~\ref{fig:plot0BSR} (a)].
An analytical expression for $V_{\mrm{nl}}$ can be found in Appendix ~\ref{app:vnl}. 
If there is a spin current flowing in the channel, a nonzero $V_{\mrm{nl}}$ will be detected since there is spin accumulation near the spin-selective detector QPC. It is assumed that the spin current is completely relaxed before reaching the equilibrium reservoir which is located far to the right of the detector QPC.  
Thus if the spin current relaxes (resonance condition) before reaching the detector QPC, no spin accumulation occurs and $V_{\mrm{nl}}$ drops.
Our numerically calculated $V_{\mrm{nl}}$ shows a plateau $B^{\perp} \approx 6-8$ T, in agreement with experimental observation\cite{Bsr}. 
The presence of a plateau, and not a sharp dip, at resonance in Fig.~\ref{fig:plot1BSR} arises as a consequence of injection in higher subbands (lower Fermi velocities) and small-angle scattering. 
After the injection in higher subbands, it is unlikely that an electron will undergo a backscattering event due to small-angle scattering \cite{SmallAngle}. Mostly, electrons will be redistributed in adjacent subbands relative to injection subband $n_{j}$. 
This redistribution of electrons among subbands with distinct resonance conditions 
manifests on $V_{\mrm{nl}}$ as a wide plateau 
depending on the relation between the distance between the QPC injector and the QPC detector $x_{\mrm{id}}$ and the spin relaxation length $\lambda_{\mrm{SR}}$ (see Appendix ~\ref{app:vnl} for further discussions).
We believe the discrepancy between our calculated $V_{\mrm{nl}}$ and the measured one away from the plateau is possibly due to additional scattering mechanisms not included in our simulations. This is a point that deserves further investigation.

Notice that in the special case where the Rashba and Dresselhaus coupling are tuned to have equal strengths $\alpha=\beta$, in the absence of cubic corrections, the effective SO magnetic field has a fixed direction in space and the DP and Elliot-Yafet mechanisms are suppressed \cite{Egues03}.

\section{\label{sec:anomalousbsr} Anomalous ballistic spin resonance}

\begin{figure}[htb]
\includegraphics[width=.47\textwidth]{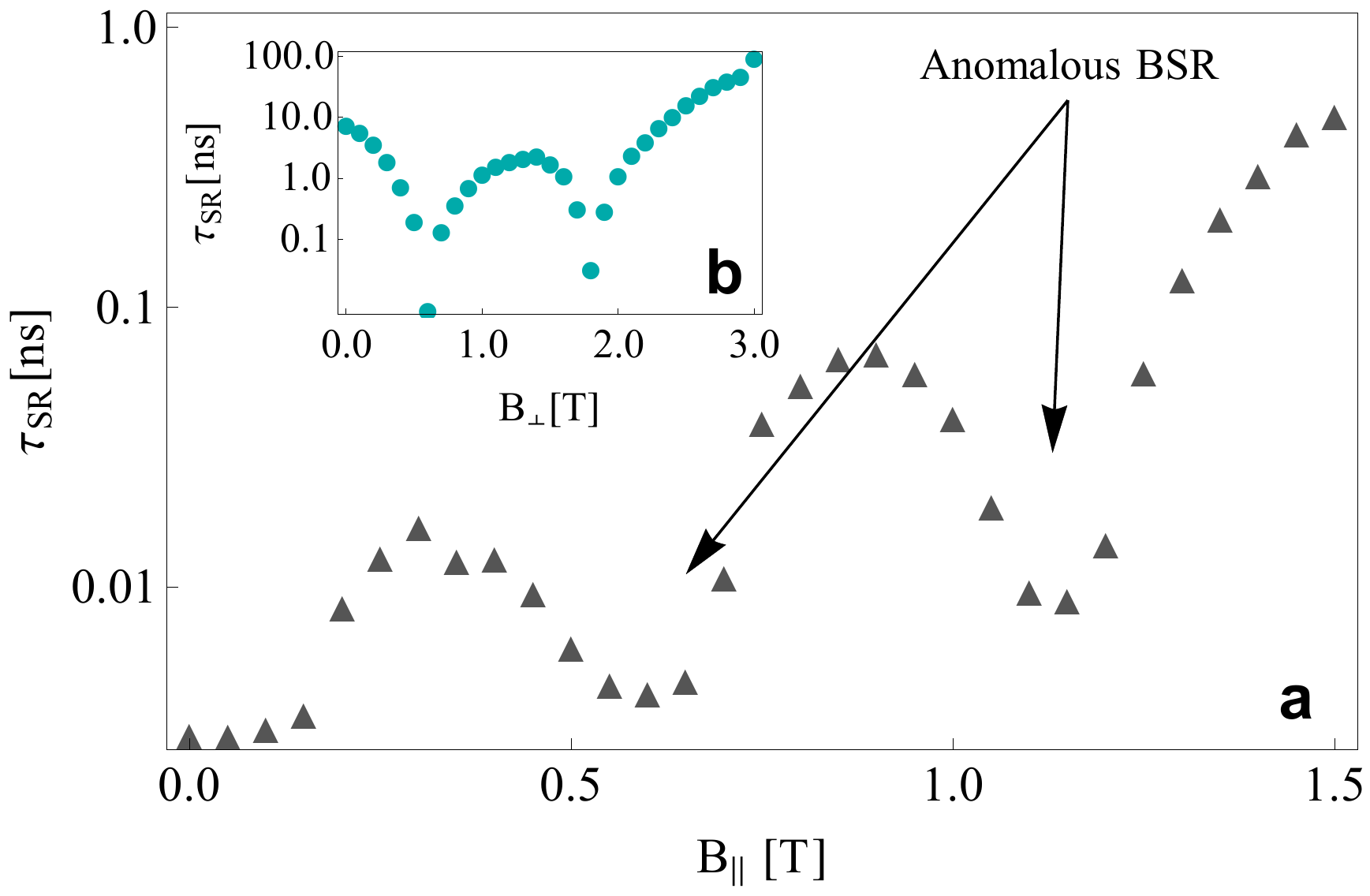}  
\caption{(a) Prediction of the dependence of the spin relaxation time $\tau_{\mrm{SR}}$ on the external magnetic field $B^{\parallel}$ in the strong SO coupling regime. In this regime, $\tau_{\mrm{SR}} (B^{\parallel})$ also presents a nonmonotonic behavior.
Anomalous BSR dips occurs around $B^{\parallel} \approx 0.6$ T with $\Lambda=1$ and $B^{\parallel} \approx 1.3$ T with $\Lambda=3$  (see arrows).
(b) $\tau_{\mrm{SR}}$ vs. $B^{\perp}$ with the resonance conditions given by $B^{\perp} \approx 0.6$ T with $\Lambda=1$ and $B^{\perp} \approx 1.9$ T with $\Lambda=3$.
Here we use the same parameters for the numerical simulation as those for GaAs wells in Sec.~\ref{sec:bsr}. 
For InAs we have that $|(\alpha + \beta)|=2$ $\mrm{meV.nm}$, $|(\alpha - \beta)|=5$ $\mrm{meV.nm}$\cite{InAs}, $\left|g\right|=14.94$\cite{winklersbook} and $m=0.026$ $m_0$\cite{vurga}.}
\label{fig:plot2BSR}
\end{figure}

In the weak SO coupling regime, the effective SO magnetic field $\varH_{\mrm{SO}}^{\perp}$ can be neglected in comparison with the $\mathbf{B}_{\mrm{ext}}$ for large fields ($|\mathbf{B}_{\mrm{ext}}| > 0.5$ T). However, in the strong SO coupling regime ($|\mathbf{B}_{\mrm{SO}}| \sim 0.3$ T for the higher subbands in InAs), this is no longer true and as a consequence we find a nonmonotonic behavior also for $\tau_{\mrm{SR}}(B^{\parallel})$. We called this emergence of extra resonance dips ``anomalous BSR''. 
Let us now analyze the cases for different orientation of an in-plane magnetic field for the strong SO regime. 
For $B^{\perp}$, the strong SO term $\varH_{\mrm{SO}}^{\perp}$ only changes substantially the resonance condition, as can be checked in Eq.~\ref{eq:resonance}. The subband-spin mixing term $\varH_{\mrm{SO}}^{\parallel}$ still acts flipping the electron spin and also quickening the spin relaxation. 
For $B^{\parallel}$, the term $\varH_{\mrm{SO}}^{\perp}$ also modifies the resonance condition. 
Moreover, this component of the SO magnetic field perpendicular to the wire can also tilt the spin initially oriented along $B^{\parallel}$
parallel to a new direction denoted by $\hat{u}$. 
Therefore, the spin-orbit induced admixture of state with opposite spins allows for the transition
$\bra{n_{j},k,\downarrow_u} \varH_{\mrm{SO}}^{\parallel} \propto \partial_y \sigma_x \ket{n_{j} \pm \Lambda,k,\uparrow_u} \neq 0$, where $\Lambda=1,3,5...$.
Thus considering the energy spectrum of the Hamiltonian  $\left[ \varH_0 + \varH_{\mrm{SO}}^{\perp} + \varH_Z^{\parallel}  \right]\ket{n,k,s_x} = \epsilon_{n,k,s_x} \ket{n,k,s_x}$, the condition for the crossing of energy levels $\epsilon_{n_j,k,\downarrow_x} =\epsilon_{n_j\pm\Lambda,k,\uparrow_x}$ occurring for the $B^{\parallel}_{\mrm{BSR}}$ is fulfilled whenever 

\begin{eqnarray}
\frac{1}{2} g \mu_B  B^{\parallel}_{\mrm{BSR}}  = \sqrt{   \left(   \frac{\pi^2 \hbar^2}{4 m L^2} \left[ \pm 2n_{j} \Lambda + \Lambda^2    \right]   \right)^2    -     \left(   (\alpha + \beta) k_F   \right)^2 }. 
\label{eq:resonance2}
\end{eqnarray}
It leads to an enhancement of the DP spin relaxation giving rise to BSR dips even when the external magnetic field is applied parallel to the quantum wire as shown in Fig.~\ref{fig:plot2BSR}.
Since this effect is enhanced in systems with a strong SO coupling strength, we choose an InAs quantum well\cite{InAs} in order to simulate and analyze the features of the anomalous BSR. Such materials contrast with GaAs where the effect is too weak to be possibly observed experimentally.
Besides, the gyromagnetic factor in InAs ($|g|=14.9$) is much larger than in GaAs ($|g|=0.44$) reducing the value of the external magnetic field $\mathbf{B}_{\mrm{BSR}}$ given by Eq.~\ref{eq:resonance}. This feature in InAs also allows us to observe higher harmonics ($\Lambda=3,5...$) even at low magnetic fields (see Fig.~\ref{fig:plot2BSR}).
A square wire confinement considered in our model was a choice motivated by the experimental observation of higher BSR dips in Ref.~\onlinecite{Bsr}.
The harmonic confinement only captures the first resonant dip, for $\Lambda=1$ as demonstrated in Appendix ~\ref{app:harmonic}.

\section{\label{sec:conclusion} Conclusion}

We study the magnetic field dependence of the spin relaxation time in multisubband quantum wires. To this end, we have developed a numerical model to take into account the DP spin relaxation mechanism in the calculation of the time-dependent spin operators. Averaging the spin dynamics over an ensemble allows us to extract the spin relaxation time $\tau_{\mrm{SR}}$ as a function of $\mathbf{B}_{\mrm{ext}}$.
We have obtained a nonmonotonic behavior for $\tau_{\mrm{SR}}$ when the external magnetic field is applied perpendicular $B^{\perp}$ to the quantum wire, which characterizes the BSR found experimentally in Ref.~\onlinecite{Bsr}. 
Within our description, BSR arises as an interplay between the DP spin relaxation mechanism and a rapid increase of the spin relaxation rate near the spin-orbit induced energy anticrossings of the quantum wire subbands with opposite spins. 
Different subbands with their distinct resonance conditions lead to an enlargement of the BSR dip into a wide plateau, in agreement with the experimental observation\cite{Bsr}.
In systems with a weak SO coupling, $\tau_{\mrm{SR}}$ varies monotonically with the external magnetic field pointing parallelly $B^{\parallel}$ to the quantum wire.
 
Nevertheless, we have also predicted a nonmonotonic behavior for $\tau_{\mrm{SR}} (B_{\parallel})$ as a consequence of the admixture of opposite spins along $\hat{x}$ due to the presence of a strong SO magnetic field $B^{\perp}_{\mrm{SO}}$. We suggest that these anomalous BSR dips can be measured in systems with strong SO coupling \cite{Sch1, Sch2}, such as an InAs quantum well.
We emphasize that our numerical model could be used to analyze the recent experimental applications of the BSR \cite{FrolovSpinTransistor, FrolovSOAnisotropy}. 
One of these applications is a new paradigm for a spin transistor.  
In this proposal, a 
gate voltage on top of the channel 
can control the enhancement or suppression of the spin relaxation time.  
Small changes in this gate voltage can modify the electronic density, Fermi velocity, and Rashba SO coupling strength. As a consequence, the BSR can be turned on and off by purely electrical means. 
Moreover, spin-orbit anisotropy was measured using BSR in a GaAs quantum well\cite{FrolovSOAnisotropy}. 
This anisotropy, which arises due to the interplay between the Rashba and Dresselhaus SO coupling strengths, could be estimated comparing the spin relaxation time for two distinct channel orientations. 
Finally, our model could also be used to study the anisotropy of the spin relaxation time \cite{Ohno} and its
dependence on the width of the wire. \cite{Chang09, Aws06, Kettemann07, Ando08, Wenk, Wu}, even in the limit of a few-subband quantum wire when the semiclassical approximation is no longer valid.

\begin{acknowledgments}
We wish to acknowledge useful discussions with J. A. Folk, S. L\"{u}scher, S. Frolov. 
This work was supported by the Brazilian agencies CNPq, Capes, FAPESP, and PRP/USP within the Research Support Center Initiative (NAP Q-NANO). 
It also received support from CIAM program (NSERC-CNPq-CONICET).
Recently, we became aware of the work in Ref.~\onlinecite{Flatte} that also investigates ballistic spin resonance in quasi-one-dimensional channels using a different approach as compared to ours.
\end{acknowledgments}

\appendix

\section{\label{app:simplemodel}DP mechanism in a quantum wire with two subbands}

In this appendix, we consider a special case of the generalized model developed in Sec.~\ref{sec:model}. Within this simplified model for a quantum wire with two subbands, the time-evolution operator can be obtained analytically and a more intuitive picture emerges for the spin relaxation in quantum wires. 

Consider the Hamiltonian given by Eq.~(\ref{eq:Hsquare}) written in the basis composed with two subband-spin Hilbert space 
$\varF = \left\{ \ket{nks}; n={1,2},k, s_y=\uparrow_y,\downarrow_y  \right\}$. 
Dividing this truncated Hilbert space in two independent subspaces $\varF_{\lambda=+} = \left\{\ket{1,k,\uparrow_y},\ket{2,k,\downarrow_y} \right\}$ and  $\varF_{\lambda=-} = \left\{ \ket{1,k,\downarrow_y},\ket{2,k,\uparrow_y} \right\}$, the Hamiltonian reads 

\begin{equation}
  \varH_\lambda
  =
  \epsilon_+ \openone
  + 
   \begin{bmatrix}
      \epsilon_- - \lambda (\alpha + \beta) k_x  & -\lambda i \alpha (p_y)_{12} / \hbar
    \\
     \lambda i \alpha (p_y)_{12} / \hbar & \epsilon_- - \lambda (\alpha + \beta) k_x  
  \end{bmatrix},
\end{equation}
where $\lambda=\pm$ denotes each subspace, $\epsilon_{\pm}= (\epsilon_1 \pm \epsilon_2) /2$ for the $\epsilon_i$ labeling the $i$th subband in the quantum wire, and the matrix element $(p_y)_{12}=\bra{1} p_y \ket{2}$.
Notice that the basis was truncated up to the second subband which still allows for inter-subband transitions. Henceforth, 
the external magnetic field was set to zero since it can cause spin relaxation by itself, even without considering the inter-subband transitions.
To show that the inter-subband transitions are responsible for the DP mechanism in quantum wires, 
it is equivalent to prove that the time-evolution operator for different paths does not constitute a set of commuting operators. As a consequence, the electron spin will precess differently for each path determined by the series of random multiple scatterings. In another words, the expectation value of the spin components for each electron in the ensemble after a time $\tau_{\mrm{SR}}$, calculated via Eq.~(\ref{eq:avg}), will correspond to random spin orientations in the Bloch sphere.  

For the sake of simplicity, we will choose a path such that the electron will move forward a distance $\Delta$ with the wavevector $+k$, undergo an elastic scattering, and then move backward the same distance with the wave vector $-k$.
So, starting with evaluating the time-evolution operator written in the basis $\varF$,

\begin{equation}
\varU(k)=exp \left[-(i/\hbar)\varH_{\lambda}(k) (\Delta/v_{F}^{j})\right]=  \begin{bmatrix}
    \Gamma^{+}(k) & 0 
    \\
    0 & \Gamma^{-}(k)
  \end{bmatrix},
\end{equation}
with $v_{F}^{j}$ the Fermi velocity considering the injection in the $j$th subband, $\Gamma^{\lambda}(k)=exp \left[-(i/\hbar)\epsilon_+  (\Delta/v_{F}^{j})\right] \times exp \left[ -(i/\hbar) \bm{\hat{n}}^{\lambda} \cdot \bm{\sigma} \left|\bm{\xi}^{\lambda}\right|  (\Delta/v_{F}^{j})\right]$ 
for $\bm{\hat{n}}^{\lambda}=\bm{\xi}^{\lambda}/ \left|\bm{\xi}^{\lambda}\right| $, where

\begin{equation}
\bm{\xi}^{\lambda}(k)=\left( 0, \xi_y^{\lambda}, \xi_z^{\lambda}(k)   \right)=
\left( 0, \lambda \frac{1}{\hbar} (\alpha - \beta) (p_y)_{12} , (\epsilon_- - \lambda (\alpha + \beta) k_x)   \right).
\end{equation}
To prove that $\left[\varU(k),\varU(-k)\right] \neq 0$ is equivalent finding that $\left[\Gamma^{\lambda}(k),\Gamma^{\lambda}(-k)\right] \neq 0$. Calculating then the latter commutator, we obtain the expression $\xi_y^{\lambda} \left[ \xi_z^{\lambda}(k) - \xi_z^{\lambda}(-k)  \right]$ which is different from zero since $\xi_y^{\lambda} \neq 0$. 
Therefore, the non-commutativity of the time-evolution operator emerges as a result of allowing inter-subband transitions causing the spin relaxation in a multisubband quantum wire.
In another words, an ensemble of initially spin-polarized electrons going through multiple scattering in a quantum wire will have their spins orientations randomized after reaching the same final destination.

Taking the limit of a strictly one-dimensional quantum wire, elementary rotation due to the SO effective magnetic field are performed around a single axis since no intersubband transitions are allowed; i.e., $\xi_y^{\lambda}=0$, consequently  $\left[\varU(k),\varU(-k)\right]=0$. Therefore, the expectation value of the spin components Eq.~(\ref{eq:avg}) for each electron will be exactly the same and dependent on the net path in the quantum wire.   
In this limit, the spin relaxation due to the DP mechanism no longer takes place in this system. \\
 
\section{\label{app:harmonic}Discussion about the harmonic confinement model}

Throughout the paper, we have used a square wire confinement in order develop a model to describe the BSR effect. Another option would be the harmonic confinement; however, we will show that this model does not capture the higher resonance dips in the spin relaxation time. 
Consider then, the electrostatic potential $V(y)$ modeled by the harmonic confinement,   

\begin{eqnarray}
\varH_h=\frac{{\bf p}^2}{2m}+\frac{1}{2}g \mu_B \left(  \mathbf{B}_{\mrm{SO}} +   \mathbf{B}_{\mrm{ext}}  \right) \cdot \boldsymbol\sigma  + \frac{1}{2} m \omega^2 y^2, 
\label{eq:Hharmonic}
\end{eqnarray}
where $\omega$ is the confinement frequency. 
Using the truncated subband-spin Hilbert space $\varF = \left\{ \ket{nks}; n={1,2,...n_{T}},k, s=\uparrow,\downarrow  \right\}$ as a basis to write $\varH_h$, in this basis we have

\begin{eqnarray}
\varH_h=\hbar \omega \left( a^{\dagger} a   + \frac{1}{2}\right)  + \frac{\hbar ^2 k^2}{2m}+ \frac{1}{2}g \mu_B \mathbf{B}_{\mrm{ext}} \cdot \boldsymbol\sigma  \nonumber\\ 
- (\alpha + \beta) k \sigma_y + i (\alpha - \beta) \sqrt{\frac{m \omega}{2 \hbar}} \left(a^{\dagger} - a \right) \sigma_x ,
\label{eq:Hharmonicladder}
\end{eqnarray}
where the creation and annihilation are given by $a^{\dagger}\ket{n}=\sqrt{n+1}\ket{n+1}$ and $a\ket{n}=\sqrt{n}\ket{n-1}$, respectively. The operator which mixes the spin and orbital states is identified as $\varH_{\mrm{SO}}^{\parallel} \propto \left(a^{\dagger} - a \right) \sigma_x$.
As we have pointed out in Sec.~\ref{sec:model} for the weak SO coupling regime, the spin resonance is absent when the external magnetic field is pointing along the quantum wire, $B^{\parallel}$. As a result, the mixing operator $\varH_{\mrm{SO}}^{\parallel}$ is not able to flip the electron spin since it is pointing along the $x$ direction.
On the other hand, the spin resonance is achieved for an external magnetic field perpendicular to the quantum wire, $B^{\perp}_{\mrm{BSR}}$, as long as the following condition is fulfilled, 

\begin{eqnarray}
\frac{1}{2} g \mu_B B^{\perp}_{\mrm{BSR}}= \pm \hbar \omega \Lambda + (\alpha + \beta) k .
\label{eq:resonanceharmonic}
\end{eqnarray}
where $\Lambda=1$. Therefore, the harmonic confinement model does not capture the higher resonance dips ($\Lambda=3,5...$) as the $B^{\perp}$ varies. 
This contrasts with the square wire confinement model which has $\Lambda=1,3,5...$, as explained in Sec.~\ref{sec:model}. 
We emphasize that these higher resonances $\Lambda=3,5...$ in the square wire confinement are distinct from the anomalous case predicted in systems with strong SO coupling.
The emergence of additional resonances in the anomalous BSR occurs due to the interplay of $\mathbf{B}_{\mrm{ext}}$ and $\mathbf{B}_{\mrm{SO}}$ even when the external magnetic field is aligned with the channel, as explained in Sec.~\ref{sec:anomalousbsr}. 

\begin{figure}
\includegraphics[width=.47\textwidth]{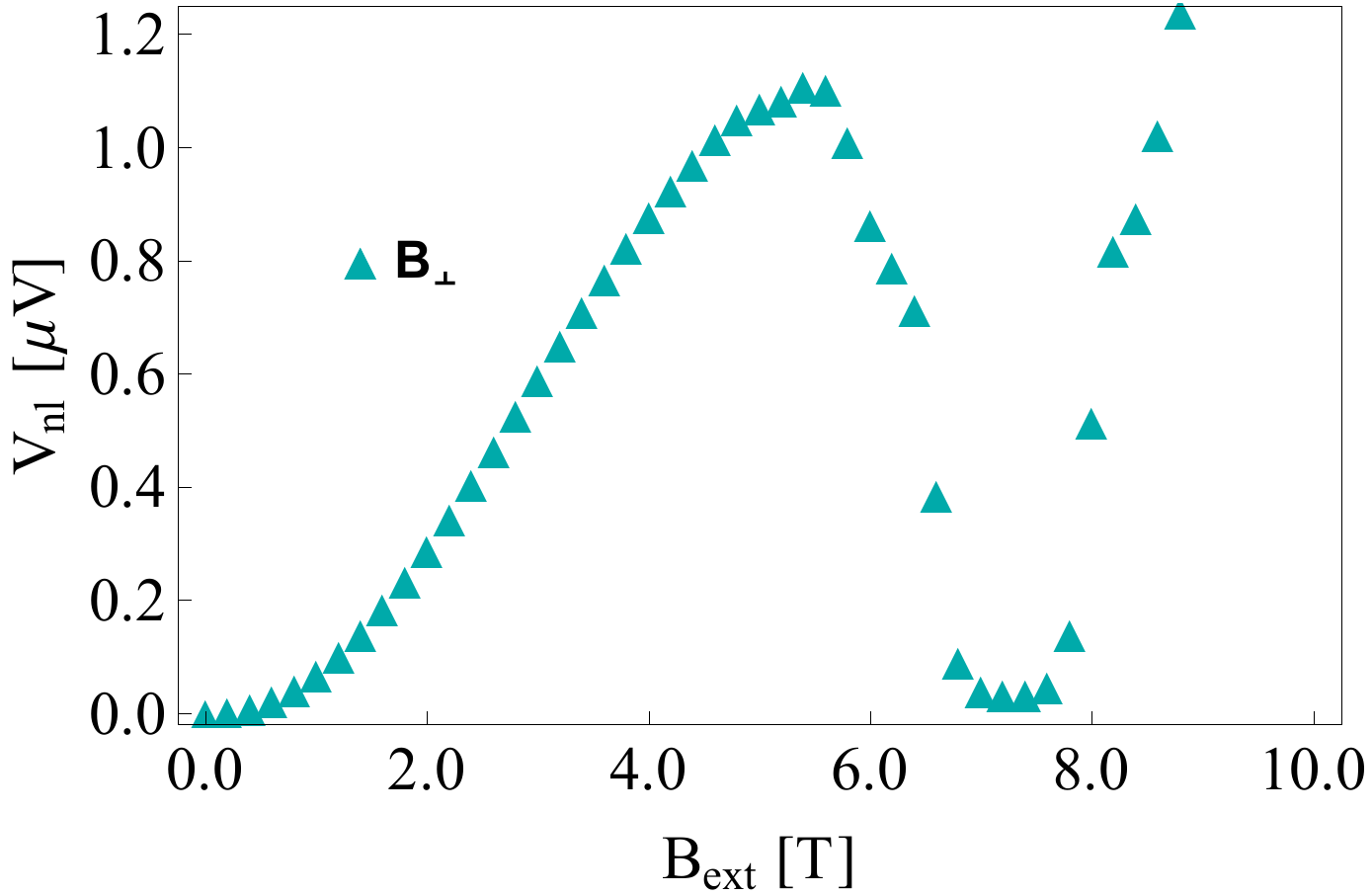}
\caption{Dependence of the nonlocal voltage $V_{\mrm{nl}}$ on the external magnetic field $B^{\perp}$ for a shorter distance between the QPC injector and the QPC detector $x_{\mrm{id}} = 5$ $\mu m$. 
All the other parameters were chosen to be the same as used in Fig.~\ref{fig:plot1BSR}. }
\label{fig:plot3BSR}
\end{figure}

\section{\label{app:vnl}Equation for the nonlocal voltage}

The nonlocal voltage $V_{\mrm{nl}}$ was derived in Ref.~\onlinecite{Frolov09} using a one-dimensional diffusion equation \cite{Jedema, Kha05}.
The explicit expression for $V_{\mrm{nl}}$ is 

\begin{equation}
V_{\mrm{nl}} =  \frac{\rho \frac{\lambda_{\mrm{SR}}}{L} I_{\mrm{inj}} P_{\mrm{inj}} P_{\mrm{det}} \sinh\left( \frac{L_r - x_{\mrm{id}}}{\lambda_{\mrm{SR}}}  \right)    }{\sinh \left( L_r/\lambda_{\mrm{SR}}  \right)   \left(    \coth\left( L_r/\lambda_{\mrm{SR}}  \right)  +   \coth\left( L_l/\lambda_{\mrm{SR}}  \right)    \right)    }
\label{eq:Vnl}
\end{equation}
where $\rho$ is the channel resistivity and $L_r$, $L_l$ denote the distance between the QPC injector and the right and left ends of the channel, respectively. The distance between the QPC injector and QPC detector is denoted by $x_{\mrm{id}}$. 
The injection current $I_{\mrm{inj}} = G_{\mrm{inj}} V_{\mrm{inj}}$, where $V_{\mrm{inj}}$ is the voltage applied across the QPC injector.  
$P_{\mrm{inj}}$ ($P_{\mrm{det}}$) denotes the spin polarization $P = \left( G_{\uparrow} - G_{\downarrow} \right) / \left( G_{\uparrow} + G_{\downarrow} \right)$ of the QPC injector (QPC detector) with the spin quantization axis defined by $\mathbf{B}_{\mrm{ext}}$. A fully polarized transmission $P \sim 1$ corresponds to a single occupied spin state, i.e., $G_{\uparrow} \sim e^2/h$ and $G_{\downarrow} \sim 0$. 
To obtain this expression for $V_{\mrm{nl}}$, a general solution to the chemical potential $\mu_{\uparrow}$, $\mu_{\downarrow}$ was found in each region of the experimental setup \cite{Jedema}    
via the one-dimensional diffusion equation $D \partial^2 V_{\mrm{nl}} / \partial x^2 = V_{\mrm{nl}} / \lambda_{\mrm{SR}}^2$.
Here the spin relaxation length $\lambda_{\mrm{SR}} = \sqrt{D  \tau_{\mrm{SR}}}$, where $D$ is the diffusion constant \cite{Jedema}.
The boundary conditions required an equilibrium spin polarization at the left and right ends of the channel, i.e., $V_{\mrm{nl}}(L_l) = V_{\mrm{nl}}(L_r) = 0$. 
Also, it was considered the  
continuity of the chemical potential 
and conservation of the spin currents across each region of the setup \cite{Jedema}.
Finally, the difference between the chemical potentials in the QPC detector and reservoir regions was calculated 
which finally results in Eq.~(\ref{eq:Vnl}), as shown by Ref.~\onlinecite{Frolov09}.
Notice that the emergence of a wide plateau in $V_{\mrm{nl}}(B^{\perp})$ depends on the distance between the QPC injector and the QPC detector $x_{\mrm{id}}$.  
This dependence can be understood comparing $x_{\mrm{id}}$ with the spin relaxation length $\lambda_{\mrm{SR}} = \sqrt{D  \tau_{\mrm{SR}}}$, where $D=v_F^2 \tau/2$ is the diffusion constant. At resonance, $\lambda_{\mrm{SR}} \sim$ $\mu m$ for the magnetic field interval $6.5-7.8$ T (determinated by the values of $n_j$ and $\Lambda$ that fulfills the resonant condition Eq.~(\ref{eq:resonance})), which is much shorter than $x_{\mrm{id}} = 20$ $\mu m$ used in the experimental setup \cite{Bsr}. 
As a consequence, the initially spin-polarized ensemble relax before reaching the QPC detector 
and $V_{\mrm{nl}}$ signal drops to zero. 
A narrower plateau can be obtained for a shorter $x_{\mrm{id}}$ comparable to $\lambda_{\mrm{SR}}$ \cite{FrolovSOAnisotropy}, as shown in Fig.~\ref{fig:plot3BSR}. 
While the nonlocal voltage plateau observed in Ref. \cite{Bsr} can be attributed to undetectable
spin accumulation near the detector, we emphasize that our numerical simulation gives a wide plateau 
for the parameters extracted from the experimental work \cite{Bsr}.


\begin{thebibliography}{99}
\bibitem{cubicDress} 
   We neglect the cubic Dresselhaus SO in our model since it does not affect strongly the magnetic field dependence of the spin relaxation time for realistic parameters and SO coupling strenghts. (see Refs.~\onlinecite{Luscher10, Jancu05}).
\bibitem{eigendecomp} We have a square matrix represented by the time-evolution operator $\varU(k,t)=\mrm{exp} \left[-(i/\hbar)\varH(k) t\right]$ in the truncated subband-spin basis $\varF$ with eigenvectors denoted by $\Lambda_1, \Lambda_2, \cdots, \Lambda_{2 n_{T}}$ and with respective eigenvalues $\lambda_1,\lambda_2, \cdots, \lambda_{2 n_{T}}$. Defining $\mathsf{P}_k \equiv \begin{bmatrix} \Lambda_1 & \Lambda_2 & \cdots & \Lambda_{2 n_{T}} \end{bmatrix}$, $\varU(k,t)$ can be decomposed via a transformation given by $\mathsf{P}_k \tilde{\varU}(k,t) \mathsf{P}_k^{-1}$, where $\tilde{\varU}(k,t)$ has only diagonal components 
$\lambda_{j} (t)=\mrm{exp}\left[-(i/\hbar)\epsilon_j t\right]$ with $\epsilon_j$ denoting the eigenenergies of the Hamiltonian (\ref{eq:Hsquare}).
\bibitem{SW}
It is time-demanding computationally to handle $2 n_T \times 2 n_T$ matrices given by $\mathsf{P}_{\gamma_\nu k_{\mrm{n_{\nu}}}}$ and $\mathsf{P}_{\gamma_\nu k_{\mrm{n_{\nu}}}}^{-1}$ derived from our Hamiltonian $\varH$.
In order to calculate these matrices, we have to restrict ourselves to a subspace of $\varF$. To do so, we perform a Schrieffer-Wolff transformation in order to isolate a subspace considering four subbands (size $n_{\mrm{sub}} = 8$) considering the injection subband $n_j$ as the lowest one in this subspace.
\bibitem{footnote1} After an injection via a spin-polarized QPC, the two main mechanisms responsible to redistributing the electrons around the Fermi energy for all quantum wire subbands are the electron-phonon interaction and the electron-electron interaction. Nonetheless, the BSR experiment (see Ref.~\onlinecite{Bsr}) was performed in the low-temperature regime ($<$ 1 K) such that the interaction causing electronic subband redistribution can be neglected. This argument relies on the dependence with temperature of the electron-phonon scattering rate $1/\tau_{e-ph} \propto T^5$ and the electron-electron scattering rate $1/\tau_{ee} \propto T^2/\epsilon_F$, where $\epsilon_F$ is the Fermi energy (see Ref.~\onlinecite{Kettemann10}).
We also neglect Elliot-Yafet mechanism since it has a small contribution to the spin relaxation rate in the ballistic regime [\onlinecite{Elliot, Yafet}].
\end{thebibliography}
\end{document}